\documentclass[submission,copyright,creativecommons]{eptcs}
\providecommand{\event}{FTSCS 2012} 
\usepackage{breakurl}             
\usepackage{llncsdoc}
\usepackage{stmaryrd}
\usepackage{mathbbold}
\usepackage{amssymb}
\usepackage{booktabs}
\usepackage{amsmath}
\usepackage{amsfonts}
\usepackage{colortbl}
\usepackage{graphicx}

\newcommand{\NotDownarrow}{\Downarrow\hspace{-0.06in}/}
\newcommand{\NotDownarrowTwo}{\Downarrow\hspace{-0.12in}/}
\newcommand{\ThreeQuotersLeft}{[\![\![}
\newcommand{\ThreeQuotersRight}{]\!]\!]}

\definecolor{mygray}{gray}{.9}

\title{A Timed Calculus for Mobile Ad Hoc Networks}
\author{Mengying Wang
\institute{Software Engineering Institute\\
East China Normal University\\
Shanghai, China} \email{mywang@sei.ecnu.edu.cn} \and Yang Lu
\institute{Department of Computer Science\\Shanghai Jiaotong
University\\Shanghai, China} \email{luyang0415@sjtu.edu.cn} }
\def\titlerunning{A Timed Calculus for Mobile Ad Hoc Networks}
\def\authorrunning{M. Wang \& Y. Lu}
\begin{document}
\maketitle

\begin{abstract}
We develop a timed calculus for Mobile Ad Hoc Networks embodying the
peculiarities of local broadcast, node mobility and communication
interference. We present a Reduction Semantics and a Labelled
Transition Semantics and prove the equivalence between them. We then
apply our calculus to model and study some MAC-layer protocols with
special emphasis on node mobility and communication interference.\\
A main purpose of the semantics is to describe the various forms of
interference while nodes change their locations in the network. Such
interference only occurs when a node is simultaneously reached by
more than one ongoing transmission over the same channel.
\end{abstract}

\section{Introduction}
\hspace*{4ex}Mobile ad hoc networks (MANETs) are complex distributed
systems that consist of a collection of wireless mobile nodes that
can dynamically self-organize into arbitrary network topologies, so
as to allow people and devices to seamlessly interwork in areas
without pre-existing communication infrastructures \cite{MAHN}.
Owing to the flexibility and convenience, their applications have
been extended from traditional military domain to a variety of
commercial areas, e.g., ambient intelligence \cite{AI}, personal
area networks \cite{PAN} and location-based services \cite{LBS}.

Wireless nodes use radio frequency channels to broadcast messages.
Compared to the conventional wired-based broadcasts like Ethernet
networks, this form of broadcast has some special features. First,
broadcasting is \emph{local}, i.e., a transmission covers only a
limited area, called a \emph{cell}, and hence reaches a (possibly
empty) subset of the nodes in the network. Second, channels are
\emph{half-duplex}: on a given channel, a node can either transmit
or receive, but cannot do both simultaneously. As a result,
communication interference can only be detected at the destination.
Further, nodes in MANETs can move arbitrarily, which makes the
network easily suffer from interference. Since interference plays an
important role in evaluating the performance of a network, it
becomes a delicate aspect of MANETs that is handled by a great
quantity of protocols (e.g., MACA/R-T\cite{MACA/R-T}).

Over the last two decades, a number of process calculi have been
proposed to model MANETs
\cite{CBSxing,CMAN1,RBPT,CNT,CMAN2,CMN,CWS,w-calculus,TCWS,CSDT}.
These calculi can be divided into two categories according to their
attentions to the network. The first group contains CBS$\#$
\cite{CBSxing}, CMAN \cite{CMAN1,CMAN2}, RBPT \cite{RBPT}, CNT
\cite{CNT}, CMN \cite{CMN}, $\omega$-calculus \cite{w-calculus} and
CSDT \cite{CSDT}. They attempt to depict \emph{local broadcast} and
\emph{node mobility}. Take CMN as an example, each node is equipped
with a location and a radius that define the cell over which the
node can transmit. When a sender broadcasts messages, only nodes
that are within its transmission cell could receive. Furthermore,
nodes are marked mobile or stationary, and mobile nodes can change
their locations randomly. Then CWS \cite{CWS} and TCWS \cite{TCWS}
constitute the second group. They focus on \emph{local broadcast}
and \emph{communication interference}. The former abstracts the
transmission into two state change events: begin transmission and
end transmission, while the latter regards the transmission as a
time consuming procedure. To our knowledge, no calculus has
integrated all of the three peculiarities, especially including node
mobility and communication interference.

In this paper, we present a \emph{timed calculus for mobile ad hoc
networks} (TCMN), which extends CWS \cite{CWS} and deals with all of
the three issues. A central concern of our calculus is to describe
the forms of interference while nodes move their locations in the
network. Towards local broadcast, we write $\mathsf{n[Q]^{c}_{l,r}}$
to stand for a node identified by $\mathsf{n}$, located at
$\mathsf{l}$, executing process $\mathsf{Q}$, and which can transmit
data over channel $\mathsf{c}$ in a cell centered at $\mathsf{l}$
with radius $\mathsf{r}$. As for node mobility, measures vary
according to the specific situation. For instance, nodes that
presently participate in no transmission could move arbitrarily
without any impact on the environment. However, the movement of an
active transmitter may affect the receptions of active receivers:
some may get an error or interference, since they passively leave or
enter the transmitter's transmission cell. Finally, with regard to
communication interference, we assume all wireless nodes have been
synchronized by some clock synchronization protocol
\cite{GlobalClock,TinySync}. Then we consider a transmission
proceeds in discrete steps which are represented by occurrences of a
simple action $\sigma$ to denote passing of one time unit. And if a
receiver is exposed to more than one ongoing transmission over the
same channel, it detects an interference.

In concurrency theory, Labelled Transition Semantics (LTS) is the
most popular way of giving operational semantics since the
transitions of a LTS expose the full behavior of the system (its
internal activities as well as the interactions with the
environment) which is required for defining behavioral equivalences
and providing powerful proof techniques. However, sometimes the
rules of a LTS may be difficult to understand particularly when the
calculi relates to node mobility like \cite{CMAN1,CMAN2,CMN}. Hence,
a different form of operational semantics, named Reduction Semantics
(RS), is introduced. RS only concerns the internal activities of a
system, so it is easier to grasp. Besides, RS can be used to check
the correctness of a LTS, by proving consistency with the LTS. For
these reasons, we define both RS and LTS semantics for our TCMN and
prove that they coincide.

We end this section with an outline of the paper. In Section 2, we
define the syntax of our core language. Then in Section 3, we
provide a RS for our calculus which specifies how an unbounded
number of system components can be involved in an atomic
interaction. Next a LTS that captures all the possible interactions
of a term with its environment is proposed in Section 4. The
equivalence between the RS and the LTS semantics is proved in
Section 5. In Section 6 and 7, we extend our core language by adding
some new operators to model some MAC-layer collision avoidance
protocols: CSMA and MACA/R-T. We prove that the CSMA protocol
doesn't solve the issue of node mobility while the MACA/R-T protocol
is robust against node mobility. Finally, in section 8, we summarize
our contributions and present the future work.
\section{The Core Language}
\begin{table}[!t]
\begin{center}
\begin{tabular}{>{\columncolor{mygray}\sf}c>{\columncolor{mygray}\sf}c>{\columncolor{mygray}\sf}c>{\columncolor{mygray}\sf}c}
\toprule
\rowcolor{white}\hspace*{-2.4ex} \textbf{Table 1.} The Syntax\\
\hline
\hspace*{-10ex} \emph{Networks:} & & & \\
\hspace*{-9.5ex} $\mathsf{N \overset{def}=0}$ & \hspace*{-8ex} empty network & \hspace*{3.5ex} $\mathsf{|\hspace*{1ex}n[Q]^{c}_{l,r}}$ & \hspace*{-10ex} node\\
\hspace*{-3.4ex} $\mathsf{|\hspace*{1.2ex}N|N}$ & \hspace*{-3ex} parallel composition & & \\
\hspace*{-10ex} \emph{Processes:} & & & \\
\hspace*{-9ex} $\mathsf{Q \overset{def}=P}$ & \hspace*{-4.8ex} non-active process & $\mathsf{|\hspace{1ex}A}$ & \hspace*{-1ex} active process\\
\hspace*{-9.5ex} $\mathsf{P \overset{def}=0}$ & \hspace*{-11.5ex} termination & \hspace*{6ex} $\mathsf{|\hspace{1ex}out\langle u\rangle.P}$ & \hspace*{-8ex} output\\
\hspace*{-1ex} $\mathsf{|\hspace{1ex}in(x).P}$ & \hspace*{-17.6ex} input & & \\
\hspace*{-5ex} $\mathsf{A \overset{def}=\langle v \rangle^{\delta}.P}$ & \hspace{-10ex} active output & \hspace*{3.5ex} $\mathsf{|\hspace*{1ex}(x)^{\delta}_{v}.P}$ & \hspace*{-3ex} active input\\
\hspace*{-13ex} \emph{Values:} & & & \\
\hspace*{-9.5ex} $\mathsf{u \overset{def}=\hspace*{0.6ex}x}$ & \hspace*{-15.5ex} variable & \hspace*{-1.2ex} $\mathsf{|\hspace*{1.2ex}v}$ & \hspace*{-3ex} closed value\\
\hspace*{-10ex} \emph{Functions:} & & & \\
\hspace*{-7ex} $\mathsf{f \overset{def}=\hspace*{0.6ex}\Lbag u\Rbag}$ & \hspace*{-10ex} time function & \hspace*{1ex} $\mathsf{|\hspace{1ex}\ThreeQuotersLeft u \ThreeQuotersRight}$ & \hspace*{3.8ex} evaluation function\\
\hspace*{-1ex} $\mathsf{|\hspace{1ex}d(l_{1},l_{2})}$ & \hspace*{-6.5ex} distance function & & \\
& \multicolumn{3}{>{\columncolor{mygray}\sf}c}{\hspace*{18ex}where $\delta$ is a positive integer greater than 0}\\
\bottomrule
\end{tabular}
\end{center}
\end{table}
\vspace*{-1ex}
\hspace*{4ex}In Table 1, we present the core of TCMN.
The syntax is defined in a two-level structure: a lower one for
\emph{processes} which describes the possible status of a node, and
an upper one for \emph{networks}. For easy understanding, in this
section we only focus on those operators that are necessary for
communication while the extended language will be presented in
Section 6.

Generally, we use letters $\mathsf{a...c}$ for channels,
$\mathsf{m...o}$ for identifiers, $\mathsf{x...z}$ for variables,
$\mathsf{u}$ for values that can be transmitted over channels: these
include variables and closed values, and $\mathsf{v}$ for closed
values, i.e. values that contain no variables. $\mathsf{\Lbag
u\Rbag}$ is a unary function designed to estimate the number of time
units required for the transmission of the value $\mathsf{u}$. Since
only closed values will be used in transmissions, we assume the
existence of an evaluation function $\mathsf{\ThreeQuotersLeft .
\ThreeQuotersRight}$ to return the closed form of a value. Finally,
we do not set how locations should be specified, the only assumption
is that they should be comparable, so to determine whether a node is
in or out of the transmission cell of another node. We do so by
introducing a function $\mathsf{d}$ which takes two locations as
parameters and returns the distance between them.

Networks are collections of nodes (which actually represent devices)
that run in parallel and use same channels to communicate with each
other. We use the symbol $\mathsf{0}$ to stand for the empty
network, and $\mathsf{n[Q]^{c}_{l,r}}$ to denote a node identified
by $\mathsf{n}$, located at $\mathsf{l}$, executing process
$\mathsf{Q}$, and which can transmit data over channel $\mathsf{c}$
in a cell centered at $\mathsf{l}$ with radius $\mathsf{r}$. We
write $\mathsf{N|N}$ to indicate a parallel composition of two
sub-networks $\mathsf{N}$.

Processes, living within the nodes, are sequential. For convenience,
we divide processes into two categories: \emph{non-active} and
\emph{active}. An active process is a process that is currently
transmitting or receiving data, e.g., an active output process
$\mathsf{\langle v \rangle^{\delta}.P}$ denotes a transmitting
process, and its transmission of value $\mathsf{v}$ will complete
after $\mathsf{\delta}$ time units. Similarly, an active input
process $\mathsf{(x)^{\delta}_{v}.P}$ represents a receiving
process, and its reception of value $\mathsf{v}$ will last for the
next $\mathsf{\delta}$ instants of time. In the non-active process
constructs, the symbol $\mathsf{0}$ stands for a terminated process.
$\mathsf{out\langle u\rangle.P}$ is an output process willing to
broadcast the value $\mathsf{v=\ThreeQuotersLeft u
\ThreeQuotersRight}$, and once the transmission starts, the process
evolves into the active output process $\mathsf{\langle v
\rangle^{\delta}.P}$, where $\mathsf{\delta=\Lbag u\Rbag}$ is the
time necessary to transmit the value $\mathsf{v}$.
$\mathsf{in(x).P}$ indicates an input process willing to receive
data, and when the beginning of a transmission $\mathsf{v}$ in the
following $\mathsf{\delta}$ time units is captured clearly (i.e.
without interference), the process becomes the active input process
$\mathsf{(x)^{\delta}_{v}.P}$. A node with an active output process
inside is named \emph{active transmitter}. Similarly, active input
processes and non-active processes are included separately in
\emph{active receivers} and \emph{non-active nodes}.

We assume that each node has a unique identifier, and different
nodes cannot be located at the same position at the same time. We
consider such networks \emph{well-formed}. Since nodes cannot be
created or destroyed, the well-formedness of a network is always
preserved as the network evolves. In the remainder of the paper, all
networks are well formed, and we use a number of notational
conventions. Process $\mathsf{Q}$ stands for either a non-active or
an active process while $\mathsf{P}$ and $\mathsf{A}$ represent
non-active and active processes separately. We identify
$\mathsf{\langle v \rangle^{\delta}.P=P}$ and
$\mathsf{(x)^{\delta}_{v}.P=P\{v/x\}}$ if $\mathsf{\delta=0}$. We
write $\mathsf{out\langle u\rangle}$ for $\mathsf{out\langle
u\rangle.0}$, and $\mathsf{\langle v \rangle^{\delta}}$ for
$\mathsf{\langle v \rangle^{\delta}.0}$.
\section{Reduction Semantics}
\hspace*{4ex}In this section, we study the reduction semantics (RS)
for TCMN. In the literature \cite{CWS}, the only internal activity
is a broadcast which is modelled by two events: begin transmission
event and end transmission event. Yet in our system, a new type of
internal activity: a migration is appended to depict node movement.
In our model, the broadcast will be described by a begin
transmission event and several time passing events (as shown in
Table 2), while the migration will be represented by a node movement
event from a specific node (as shown in Table 3). Among these three
types of events, the \emph{begin transmission event} (i.e. a node
initiates a transmission) has the same meaning as that in
\cite{CWS}, while the \emph{time passing event} (i.e., a unit of
time delays) is imported to replace the end transmission event in
\cite{CWS}, and the \emph{node movement event} (i.e., a node moves
from one location to another) is a newly added event.

In our RS for core TCMN, a reduction denotes either a begin
transmission event, or a time passing event, or a node movement
event. In order to handle the interaction among an unbounded number
of processes, we use rule schemas instead of simple rules to
demonstrate the reductions. Also, since a reduction, e.g. a begin
transmission event, cannot be performed inside arbitrary contexts:
one should guarantee that the current context meets the specific
conditions, the minimal information about the communication is
attached to the reduction. Further, in order to model communication
interference, we store all the needed active transmitters'
information in a global set $\mathsf{T}$ which displays in any
reduction to determine whether a node is simultaneously reached by
more than one transmission over the same channel. For this reason,
the reduction semantics is named RST: RS with parameter
$\mathsf{T}$. The component $\mathsf{T}$ is a set of triples
$\mathsf{(l,r,c)}$ with each $\mathsf{l}$,$\mathsf{r}$,$\mathsf{c}$
in a triple represents location, radius and channel of an active
transmitter separately. For simplicity, the semantics does not
automatically update the set $\mathsf{T}$. Therefore, when a
reduction is performed, the new $\mathsf{T}$ which will be used in
the next one has to be manually computed. However, it is not
difficult to modify the rules so that they also produce the new
$\mathsf{T}$.

As usual in process calculi, the reduction semantics relies on an
auxiliary relation, called \emph{structural congruence}, denoted by
$\mathsf{\equiv}$, to allow the manipulation of the term structure
so as to bring the participants of a potential interaction into
contiguous positions. Here we define a smallest congruence including
associativity, commutativity and identity over the empty
network:\\[1.5ex]
\hspace*{8ex}$\mathsf{N|(N'|N'')\equiv (N|N')|N''~~N|N'\equiv
N'|N~~N|0\equiv N}$\\[1.5ex]
Next are some useful notations that will be used in RST:
\begin{enumerate}
 \item [$\bullet$] $\mathsf{T|_{l,c}}$ is the subset of the active transmitters
 $\mathsf{T}$ whose transmissions are synchronized on channel $\mathsf{c}$ and can
 reach a node located at $\mathsf{l}$. Formally, \\[1ex]
   \hspace*{4ex}$\mathsf{T|_{l,c}=\{(l',r',c')|(l',r',c')\in T \wedge d(l',l)\leq r' \wedge
   c'=c\}}$\\[-1.5ex]
 \item [$\bullet$] $\mathsf{(l,r,c)\NotDownarrowTwo_{i} N}$ holds if Network
 $\mathsf{N}$ contains no input nodes $\mathsf{n[in(x).P]^{c}_{l',r'}}$
 or $\mathsf{n[(x)^{\delta}_{v}.P]^{c}_{l',r'}}$ for which $\mathsf{d(l,l')\leq r}$
 is true (i.e., a transmission from a node located at $\mathsf{l}$, with radius
 $\mathsf{r}$, synchronized on $\mathsf{c}$ reaches no input nodes in $\mathsf{N}$).\\[-1.5ex]
 \item [$\bullet$] $\mathsf{(l,r,c)\NotDownarrowTwo_{ai} N}$ holds if Network $\mathsf{N}$
 contains no active input nodes $\mathsf{n[(x)^{\delta}_{v}.P]^{c}_{l',r'}}$ for which
 $\mathsf{d(l,l')\leq r}$ is true (i.e., a transmission from a node located at $\mathsf{l}$, with radius
 $\mathsf{r}$, synchronized on $\mathsf{c}$ reaches no active input nodes in $\mathsf{N}$).
\end{enumerate}

\begin{table}[!t]
\begin{center}
\begin{tabular}{>{\columncolor{mygray}\sf}c}
\toprule
\rowcolor{white}\hspace*{-1.6ex}\textbf{Table 2.} Reduction Semantics - Begin transmission and time passing event\\
\hline\\[-0.5ex]
\hspace*{22ex}[RST-BEGIN]\hspace*{13ex}[RST-PASS-NULL]\\[0.5em]
\normalsize{$\mathsf{\frac{\forall h\in I\cup J\cup K.d(l,l_{h})\leq
r~~\forall i\in I.T|_{l_{i},c}=\emptyset~~\forall j \in
J.T|_{l_{j},c}\neq \emptyset}{T\rhd n[out\langle
u\rangle.P]^{c}_{l,r}|\prod\limits_{h\in I\cup
J}n_{h}[in(x_{h}).P_{h}]^{c}_{l_{h},r_{h}}|\prod\limits_{k\in
K}n_{k}[(x_{k})^{\delta_{k}}_{v_{k}}.P_{k}]^{c}_{l_{k},r_{k}}\hookrightarrow
^{c}_{l,r}}}$}~~~~\scriptsize{$\mathsf{T\rhd 0 \hookrightarrow^{\sigma} 0}$}\\[1.5em]
\scriptsize{\hspace*{-18ex}$\mathsf{n[\langle \ThreeQuotersLeft u
\ThreeQuotersRight \rangle^{\Lbag
u\Rbag}.P]^{c}_{l,r}|\prod\limits_{i\in I}n_{i}[(x_{i})^{\Lbag
u\Rbag}_{\ThreeQuotersLeft u
\ThreeQuotersRight}.P_{i}]^{c}_{l_{i},r_{i}}|\prod\limits_{j\in
J}n_{j}[in(x_{j}).P_{j}]^{c}_{l_{j},r_{j}}|\prod\limits_{k\in
K}n_{k}[P_{k}\{\bot/x_{k}\}]^{c}_{l_{k},r_{k}}}$}\\[2em]

\hspace*{16ex}[RST-SENDING]\hspace*{18ex}[RST-PASS-NA]\\[0.5em]
\normalsize{$\mathsf{\frac{\delta>0~~\forall i\in I.d(l,l_{i})\leq
r}{T\rhd n[\langle v
\rangle^{\delta}.P]^{c}_{l,r}|\prod\limits_{i\in
I}n_{i}[(x_{i})^{\delta}_{v}.P_{i}]
^{c}_{l_{i},r_{i}}\hookrightarrow^{\sigma}n[\langle v
\rangle^{\delta-1}.P]^{c}_{l,r}|\prod\limits_{i\in
I}n_{i}[(x_{i})^{\delta-1}_{v}.P_{i}]^{c}_{l_{i},r_{i}}}}$}~~~\scriptsize{$\mathsf{T\rhd n[P]^{c}_{l,r}\hookrightarrow^{\sigma} n[P]^{c}_{l,r}}$}\\[2.5em]

\hspace*{-1ex}[RST-CONT]\hspace*{5ex}[RST-CONT-PASS]\hspace*{7ex}[RST-CONGR]\\[0.5em]
\normalsize{$\mathsf{\frac{T\rhd N\hookrightarrow
^{c}_{l,r}N'~~(l,r,c)\NotDownarrow_{i} N''}{T\rhd
N|N''\hookrightarrow ^{c}_{l,r}N'|N''}}~~\mathsf{\frac{T\rhd
N\hookrightarrow^{\sigma} N'~~T\rhd N''\hookrightarrow^{\sigma}
N'''}{T\rhd
N|N''\hookrightarrow^{\sigma}N'|N'''}}~~\mathsf{\frac{N\equiv
N'~~T\rhd N'\hookrightarrow ^{\&} N''~~N''\equiv N'''}{T\rhd
N\hookrightarrow^{\&} N'''}}$}\\
\bottomrule
\end{tabular}
\end{center}
\end{table}
Let's explain the rules in Table 2 and 3. Rule RST-BEGIN is used to
derive begin transmission reduction. As in \cite{CWS}, it rewrites
atomically an output node $\mathsf{n[out\langle
u\rangle.P]^{c}_{l,r}}$ which is intending to initiate a
transmission and all the receiver nodes that are not only in its
transmission cell but also synchronized on the same channel
$\mathsf{c}$. After this reduction, the output process evolves into
$\mathsf{\langle \ThreeQuotersLeft u \ThreeQuotersRight
\rangle^{\Lbag u\Rbag}.P}$ indicating an active output process that
will transmit the evaluation result $\mathsf{\ThreeQuotersLeft u
\ThreeQuotersRight}$ of value $\mathsf{u}$ in the following
$\mathsf{\Lbag u \Rbag}$ time units. The effect of the begin
transmission event on each receiver varies according to the
structure of each receiver and the set $\mathsf{T}$. There are three
different situations, corresponding to the sets $\mathsf{I}$,
$\mathsf{J}$ and $\mathsf{K}$. Processes in $\mathsf{I}$ represent
normal inputs. Since their environments are silent
($\mathsf{T|_{l_{i},c}=\emptyset}$), they become active inputs of
the form $\mathsf{(x_{i})^{\Lbag u\Rbag}_{\ThreeQuotersLeft u
\ThreeQuotersRight}.P_{i}}$ and start receiving data
$\mathsf{\ThreeQuotersLeft u \ThreeQuotersRight}$ for the next
$\mathsf{\Lbag u\Rbag}$ time units. By contrast, for processes in
$\mathsf{J}$, as they are currently reached by at least one other
transmission ($\mathsf{T|_{l_{j},c}\neq \emptyset}$), they could not
receive the begin transmission event clearly and stay idle. Finally,
processes in $\mathsf{K}$ are active inputs, i.e., they are
receiving another transmission, so the new begin transmission event
causes interference, denoted by receiving symbol $\mathsf{\bot}$.

Rule RST-SENDING deals with the time passing event for active
processes. Initially, the active output process $\mathsf{\langle v
\rangle^{\delta}.P}$ requires $\mathsf{\delta}$ time units to
complete the date transmission. After a time interval, the remaining
time would be $\mathsf{\delta-1}$ units for both sender and
receivers. Meanwhile rule RST-PASS-NA and RST-PASS-NULL handle the
time passing event for non-active processes and empty networks
respectively. No matter how time flies, they remain unchanged.

\begin{table}[!t]
\begin{center}
\begin{tabular}{>{\columncolor{mygray}\sf}c}
\toprule
\rowcolor{white}\hspace*{-16.5ex}\textbf{Table 3.} Reduction Semantics - Node movement event\\
\hline\\[-0.5ex]
\hspace*{2ex}[RST-MOVE-AO]\\[0.5em]
\normalsize{$\mathsf{\frac{\forall i\in I.d(l,l_{i})\leq r\wedge
d(l',l_{i})\leq r~~\forall j\in J.d(l,l_{j})\leq r\wedge
d(l',l_{j})>r~~\forall k\in K.d(l,l_{k})>r\wedge d(l',l_{k})\leq
r}{T\rhd n[\langle v \rangle^{\delta}.P]^{c}_{l,r}|\prod\limits_{h
\in I\cup J}n_{h}[(x_{h})^{\delta}_{v}.P_{h}]
^{c}_{l_{h},r_{h}}|\prod\limits_{k \in
K}n_{k}[(x_{k})^{\delta_{k}}_{v_{k}}.P_{k}]
^{c}_{l_{k},r_{k}}\hookrightarrow^{c}_{l:l',r}}}$
}\\[1.5em]
\scriptsize{\hspace*{0ex}$\mathsf{n[\langle v
\rangle^{\delta}.P]^{c}_{l',r}|\prod\limits_{i \in
I}n_{i}[(x_{i})^{\delta}_{v}.P_{i}]
^{c}_{l_{i},r_{i}}|\prod\limits_{j \in
J}n_{j}[P_{j}\{\epsilon/x_{j}\}]^{c}_{l_{j},r_{j}}|\prod\limits_{k
\in K}n_{k}[P_{k}\{\bot/x_{k}\}]^{c}_{l_{k},r_{k}}}$}\\[1.5em]

\hspace*{0ex}[RST-MOVE-AI1]\\[0.5em]
\normalsize{$\mathsf{\frac{d(l,l_{i})\leq r_{i}\wedge
d(l',l_{i})>r_{i}}{T\rhd
n[(x)^{\delta}_{v}.P]^{c}_{l,r}|n_{i}[\langle v
\rangle^{\delta}.P]^{c}_{l_{i},r_{i}}\hookrightarrow
n[P\{\epsilon/x\}]^{c}_{l',r}|n_{i}[\langle v
\rangle^{\delta}.P]^{c}_{l_{i},r_{i}}}}$}\\[1.5em]

[RST-MOVE-AI2]\\[0.5ex]
\normalsize{$\mathsf{\frac{d(l,l_{i})\leq r_{i}\wedge
d(l',l_{i})\leq r_{i}~~T|_{l,c}=T|_{l',c}}{T\rhd
n[(x)^{\delta}_{v}.P]^{c}_{l,r}|n_{1}[\langle v
\rangle^{\delta}.P]^{c}_{l_{i},r_{i}}\hookrightarrow
n[(x)^{\delta}_{v}.P]^{c}_{l',r}|n_{i}[\langle v
\rangle^{\delta}.P]^{c}_{l_{i},r_{i}}}}$}\\[1.5em]

\hspace*{-2ex}[RST-MOVE-AI3]\\[0.5em]
\normalsize{$\mathsf{\frac{d(l,l_{i})\leq r_{i}\wedge
d(l',l_{i})\leq r_{i}~~\forall j\in J.d(l,l_{j})>r_{j}\wedge
d(l',l_{j})\leq r_{j}~~T|_{l',c}=T|_{l,c}\cup J}{T\rhd
n[(x)^{\delta}_{v}.P]^{c}_{l,r}|n_{i}[\langle v
\rangle^{\delta}.P]^{c}_{l_{i},r_{i}}|\prod\limits_{j \in
J}n_{j}[\langle v_{j}\rangle^{\delta_{j}}.P_{j}]
^{c}_{l_{j},r_{j}}\hookrightarrow
n[P\{\bot/x\}]^{c}_{l',r}|n_{i}[\langle v
\rangle^{\delta}.P]^{c}_{l_{i},r_{i}}|\prod\limits_{j \in
J}n_{j}[\langle v_{j}\rangle^{\delta_{j}}.P_{j}]
^{c}_{l_{j},r_{j}}}}$}\\[2em]

\hspace*{1ex}[RST-MOVE-NA]\hspace*{6ex}[RST-CONT-MOVE]\hspace*{5.5ex}[RST-CONT-INT]\\[0.5em]
\scriptsize{$\mathsf{T\rhd n[P]^{c}_{l,r}\hookrightarrow
n[P]^{c}_{l',r}}$}~~~~\normalsize{$\mathsf{\frac{T\rhd
N\hookrightarrow ^{c}_{l:l',r}N'~~(l,r,c)\NotDownarrow_{ai}
N''\wedge (l',r,c)\NotDownarrow_{ai} N''}{T\rhd N|N''\hookrightarrow
^{c}_{l:l',r}N'|N''}}~~~~\mathsf{\frac{T\rhd N\hookrightarrow
N'}{T\rhd N|N''\hookrightarrow N'|N''}}$}\\
\bottomrule
\end{tabular}
\end{center}
\end{table}
Rule RST-MOVE-AO, RST-MOVE-AI1, RST-MOVE-AI2, RST-MOVE-AI3, and
RST-MOVE-NA are all used to describe node movements. In RST-MOVE-AO,
an active transmitter moves from $\mathsf{l}$ to $\mathsf{l'}$. Then
for active receivers in set $\mathsf{I}$, as they are always
reachable no matter from $\mathsf{l}$ or $\mathsf{l'}$, they
continue to receive data normally. As for active receivers in set
$\mathsf{J}$, since they are reachable from $\mathsf{l}$ but not
from $\mathsf{l'}$, they get an error, represented by a special sign
$\mathsf{\epsilon}$. Finally, active receivers in set $\mathsf{K}$,
which are reachable from $\mathsf{l'}$ but not from $\mathsf{l}$,
are receiving another transmission, so the newly joined transmitter
will make them get interference. Rule RST-MOVE-AI1, RST-MOVE-AI2 and
RST-MOVE-AI3 depict all the different movements of an active
receiver. In RST-MOVE-AI1, the active receiver moves from
$\mathsf{l}$ to $\mathsf{l'}$ which makes the original transmission
no longer receivable, hence it gets an error. While in RST-MOVE-AI2,
although the active receiver moves from $\mathsf{l}$ to
$\mathsf{l'}$, it has always been in the transmitter's transmission
cell and there is no more active transmitter in $\mathsf{l'}$, so
the active receiver remains unchanged. On the contrary, in
RST-MOVE-AI3, when the active receiver arrives at $\mathsf{l'}$,
some other transmissions in $\mathsf{l'}$ interfere with its
original one. As a result, the active receiver obtains an
interference. Rule RST-MOVE-NA is straightforward, for non-active
nodes, their movement will not affect the environment, therefore
they can move arbitrarily without any limitations and changes.

Rule RST-CONT, RST-CONT-PASS, RST-CONT-MOVE and RST-CONT-INT are
closure rules with regard to different reduction forms
($\mathsf{\hookrightarrow ^{c}_{l,r}}$,
$\mathsf{\hookrightarrow^{\sigma}}$, $\mathsf{\hookrightarrow
^{c}_{l:l',r}}$, and $\mathsf{\hookrightarrow}$). In RST-CONT, it
provides a closure under contexts that do not contain receivers in
the transmission cell of the transmitter. Similarly, rule
RST-CONT-MOVE presents a closure under contexts that have never
contained active receivers in the transmission cell of the
transmitter when the transmitter moves from $\mathsf{l}$ to
$\mathsf{l'}$. Rule RST-CONT-INT is analogous to the previous two
except that it concerns \emph{internal events} (i.e., a node
movement event from an active receiver or a non-active node). Rule
RST-CONT-PASS is the time synchronization, it defines a closure
under contexts that are also affected by the time passing event.

The last rule, RST-CONGR is a closure rule under structural
congruence, where $\mathsf{\hookrightarrow ^{\&}}$ ranges over
$\mathsf{\hookrightarrow ^{c}_{l,r}}$,
$\mathsf{\hookrightarrow^{\sigma}}$, $\mathsf{\hookrightarrow
^{c}_{l:l',r}}$ and $\mathsf{\hookrightarrow}$ for some
$\mathsf{c}$, $\mathsf{l}$, $\mathsf{l'}$ and $\mathsf{r}$.
\section{Labelled Transition Semantics}
\begin{table}[!t]
\begin{center}
\begin{tabular}{>{\columncolor{mygray}\sf}c}
\toprule
\rowcolor{white}\hspace*{-34ex} \textbf{Table 4.} Labelled Transitions for Processes\\
\hline\\[-0.5ex]
\normalsize{$\mathsf{\frac{\ThreeQuotersLeft u \ThreeQuotersRight=v~~\Lbag u\Rbag=\delta}{out\langle u \rangle.P\xrightarrow{!v:\delta} \langle v \rangle^{\delta}.P}}$[PS-OUT$_{begin}$]~~~~$\mathsf{\frac{\delta>0}{\langle v \rangle^{\delta}.P\xrightarrow{\sigma} \langle v \rangle^{\delta-1}.P}}$[PS-OUT$_{send}$]}\\[2em]

\normalsize{$\mathsf{\frac{-}{in(x).P\xrightarrow{?v:\delta} (x)^{\delta}_{v}.P}}$[PS-IN$_{begin}$]~~~~$\mathsf{\frac{\delta>0}{(x)^{\delta}_{v}.P\xrightarrow{\sigma} (x)^{\delta-1}_{v}.P}}$[PS-IN$_{receive}$]}\\[2em]

\normalsize{$\mathsf{\frac{-}{in(x).P\xrightarrow{?\bot} in(x).P}}$[PS-IN$_{wait}$]~~~~$\mathsf{\frac{-}{(x)^{\delta}_{v}.P\xrightarrow{?\bot} P\{\bot/x\}}}$[PS-IN$_{interfere}$]}\\[2em]

\normalsize{$\mathsf{\frac{-}{(x)^{\delta}_{v}.P\xrightarrow{?\epsilon} P\{\epsilon/x\}}}$[PS-IN$_{err}$]~~~~$\mathsf{\frac{-}{P\xrightarrow{\sigma} P}}$[PS-PASS]~~~~$\mathsf{\frac{\alpha\in\{?v:\delta,?\epsilon,?\bot\}~~Q\notin IQ}{Q\xrightarrow{\alpha} Q}}$[PS-NOIN]}\\[1.5em]

\footnotesize{\hspace{27ex}where $\mathsf{IQ}$ is the set of processes of the form $\mathsf{in(x).P}$ or $\mathsf{(x)^{\delta}_{v}.P}$}\\
\bottomrule
\end{tabular}
\end{center}
\end{table}
\hspace*{4ex}We divide our Labelled Transition Semantics (LTS) into
two set of rules corresponding to the two-level structure of our
language. Table 4 contains the rules for the processes, while Table
5 and 6 presents those for the networks.

In the process semantics, a transition has the form $\mathsf{Q\xrightarrow{\alpha}Q'}$, where the grammar for $\mathsf{\alpha}$ is: \\[1ex]
\hspace*{8ex}$\mathsf{\alpha~:=~!v:\delta~|~?v:\delta~|~?\bot~|~?\epsilon~|~\sigma}$
\\[1ex]
Label $\mathsf{!v:\delta}$ represents a begin transmission event
(i.e., a transmission of value $\mathsf{v}$ in the following
$\mathsf{\delta}$ time units) is initiated by $\mathsf{Q}$ which
then evolves into $\mathsf{Q'}$; $\mathsf{?v:\delta}$ indicates a
begin transmission event reaches $\mathsf{Q}$ and makes the process
transform into $\mathsf{Q'}$; Analogously, $\mathsf{?\bot}$ and
$\mathsf{?\epsilon}$ stand for an interference or error arrives;
finally, $\mathsf{\sigma}$ means a time passing event.

Explanations for the rules in Table 4 are as follows: in
PS-OUT$_{begin}$, the output process calculates the value
$\mathsf{u}$ and initiates the transmission of the result
$\mathsf{v}$ in the next $\mathsf{\delta}$ time units; in
PS-IN$_{begin}$, the input process successfully becomes involved
with the transmission of value $\mathsf{v}$ for the next
$\mathsf{\delta}$ instants of time; in PS-OUT$_{send}$ and
PS-IN$_{receive}$, with the time passing by, the remaining
transmission time is decreasing; in PS-IN$_{wait}$, the input
process stays idle since it could not receive the begin transmission
event clearly; in PS-IN$_{interfere}$ and PS-IN$_{err}$, an active
input process encounters an interference or error in its reception,
and hence stops receiving; Rule PS-PASS shows that the non-active
process would never change as time goes by, and PS-NOIN demonstrates
that the non-input processes would never respond to the reception of
events.

Following are some useful mathematical symbols in LTS:
\begin{enumerate}
 \item [$\bullet$] \footnotesize{$\mathsf{d(l,l')\leq r'\odot d(l,l'')\leq r'=(d(l,l')\leq r'\wedge d(l,l'')\leq r')\vee (d(l,l')> r'\wedge d(l,l'')> r')}$.}\\[-1ex]
 \item [$\bullet$] \normalsize{$\mathsf{T|_{l,c}-T|_{l',c}}$ is the set of elements
 that are contained in $\mathsf{T|_{l,c}}$ but not in $\mathsf{T|_{l',c}}$.}\\[-2ex]
 \item [$\bullet$] $\mathsf{T|_{l,c}\subset T|_{l',c}}$ holds only if
 $\mathsf{T|_{l,c}}$ is a proper subset of $\mathsf{T|_{l',c}}$.\\[-1.5ex]
\end{enumerate}
\begin{table}[!t]
\begin{center}
\begin{tabular}{>{\columncolor{mygray}\sf}c}
\toprule
\rowcolor{white}\hspace*{-1.6ex}\textbf{Table 5.} Labelled Transitions for Networks - Begin transmission and time passing event\\
\hline\\[-0.5ex]
\normalsize{$\mathsf{\frac{P\xrightarrow{!v:\delta} A}{T\rhd
n[P]^{c}_{l,r}\xrightarrow{c!v:\delta[l,r]}
n[A]^{c}_{l,r}}}$[NS-OUT]~~~~$\mathsf{\frac{Q\xrightarrow{?v:\delta}
Q'~~d(l,l')\leq r'~~T|_{l,c}=\emptyset}{T\rhd
n[Q]^{c}_{l,r}\xrightarrow{c?v:\delta[l',r']}
n[Q']^{c}_{l,r}}}$[NS-IN$_{1}$]}\\[4ex]

\normalsize{$\mathsf{\frac{Q\xrightarrow{?\bot} Q'~~d(l,l')\leq
r'~~T|_{l,c}\neq\emptyset}{T\rhd
n[Q]^{c}_{l,r}\xrightarrow{c?v:\delta[l',r']}
n[Q']^{c}_{l,r}}}$[NS-IN$_{2}$]~~~~$\mathsf{\frac{d(l,l')>r'\vee
c\neq c'}{T\rhd n[Q]^{c}_{l,r}\xrightarrow{c'?v:\delta[l',r']}
n[Q]^{c}_{l,r}}}$[NS-IN$_{3}$]}\\[4ex]

\normalsize{$\mathsf{\frac{Q\overset{\sigma} \rightarrow Q'}{T\rhd
n[Q]^{c}_{l,r}\xrightarrow{\sigma}
n[Q']^{c}_{l,r}}}$[NS-PASS]~~~~$\mathsf{\frac{-}{T\rhd
0\xrightarrow{c?v:\delta[l,r]}0}}$[NS-NULL$_{in1}$]}\\[4ex]

\normalsize{$\mathsf{\frac{-}{T\rhd
0\xrightarrow{\sigma}0}}$[NS-NULL$_{pass}$]~~~~$\mathsf{\frac{T\rhd
N_{1}\xrightarrow{c?v:\delta[l,r]} N'_{1}~T\rhd
N_{2}\xrightarrow{c!v:\delta[l,r]} N'_{2}}{T\rhd
N_{1}|N_{2}\xrightarrow{c!v:\delta[l,r]}
N'_{1}|N'_{2}}}$[NS-COM]}\\[2ex]
\hspace{17ex}\scriptsize{$\mathsf{T\rhd N_{2}|N_{1}\xrightarrow{c!v:\delta[l,r]} N'_{2}|N'_{1}}$}\\[2ex]

\normalsize{$\mathsf{\frac{T\rhd N_{1}\xrightarrow{c?v:\delta[l,r]}
N'_{1}~~T\rhd N_{2}\xrightarrow{c?v:\delta[l,r]} N'_{2}}{T\rhd
N_{1}|N_{2}\xrightarrow{c?v:\delta[l,r]}
N'_{1}|N'_{2}}}$[NS-COM$_{in}$]~~~~$\mathsf{\frac{T\rhd
N_{1}\xrightarrow{\sigma} N'_{1}~~T\rhd N_{2}\xrightarrow{\sigma}
N'_{2}}{T\rhd N_{1}|N_{2}\xrightarrow{\sigma}
N'_{1}|N'_{2}}}$[NS-SYN]}\\
\bottomrule
\end{tabular}
\end{center}
\end{table}
\vspace*{-1ex}
\hspace*{4ex}In the network semantics, transitions
are of the form $\mathsf{T\rhd N \xrightarrow{\mu} N'}$ where
$\mathsf{T}$ is the same as in Section 3. Let's comment on the rules
in Table 5 and 6. Rule NS-OUT, NS-IN$_{1}$, NS-IN$_{2}$ and
NS-IN$_{3}$ concern the communication between a transmitter and its
receivers. Rule NS-OUT shows that a node initiates a transmission.
Then rule NS-IN$_{1}$ describes the behavior of a node that is
within the transmission cell and could hear the begin transmission
event clearly, whereas NS-IN$_{2}$ handles those that detect
conflicts. Rule NS-IN$_{3}$ demonstrates that a node would not react
to transmissions that are beyond its reception range or not in its
listening channel.

\begin{table}[!t]
\begin{center}
\begin{tabular}{>{\columncolor{mygray}\sf}c}
\toprule
\rowcolor{white}\hspace*{-10.2ex}\textbf{Table 6.} Labelled Transitions for Networks - Node movement event\\
\hline\\[-0.5ex]
\normalsize{$\mathsf{\frac{-}{T\rhd n[\langle v
\rangle^{\delta}.P]^{c}_{l,r}\xrightarrow{c![(l:l'),r]} n[\langle v
\rangle^{\delta}.P]^{c}_{l',r}}}$[NS-MOVE$_{ao}$]}\\[4ex]

\normalsize{$\mathsf{\frac{(Q\not\in AIQ)\vee(c\neq
c')\vee(d(l,l')\leq r'\odot d(l,l'')\leq r')}{T\rhd
n[Q]^{c}_{l,r}\xrightarrow{c'?[(l':l''),r']}
n[Q]^{c}_{l,r}}}$[NS-MOVE$_{in1}$]}\\[4ex]

\normalsize{$\mathsf{\frac{Q\in
AIQ~Q\xrightarrow{?\epsilon}Q'~d(l,l')\leq r'\wedge
d(l,l'')>r'}{T\rhd n[Q]^{c}_{l,r}\xrightarrow{c?[(l':l''),r']}
n[Q']^{c}_{l,r}}}$[NS-MOVE$_{in2}$]}\\[4ex]

\normalsize{$\mathsf{\frac{Q\in AIQ~Q\xrightarrow{?\bot}
Q'~d(l,l')>r'\wedge d(l,l'')\leq r'}{T\rhd
n[Q]^{c}_{l,r}\xrightarrow{c?[(l':l''),r']}
n[Q']^{c}_{l,r}}}$[NS-MOVE$_{in3}$]}\\[4ex]

\normalsize{$\mathsf{\frac{Q\in AIQ~~Q\xrightarrow{?\epsilon}
Q'~~T|_{l,c}-T|_{l',c} =T|_{l,c}}{T\rhd n[Q]^{c}_{l,r}\xrightarrow{}
n[Q']^{c}_{l',r}}}$[NS-MOVE$_{ai1}$]}\\[3ex]

\normalsize{$\mathsf{\frac{Q\in AIQ~~Q\xrightarrow{?\bot}
Q'~~T|_{l,c}\subset T|_{l',c}}{T\rhd n[Q]^{c}_{l,r}\xrightarrow{}
n[Q']^{c}_{l',r}}}$[NS-MOVE$_{ai2}$]~~~~$\mathsf{\frac{Q\in
AIQ~~T|_{l,c}=T|_{l',c}}{T\rhd n[Q]^{c}_{l,r}\xrightarrow{}
n[Q]^{c}_{l',r}}}$[NS-MOVE$_{ai3}$]}\\[3ex]

\normalsize{$\mathsf{\frac{-}{T\rhd n[P]^{c}_{l,r}\xrightarrow{}
n[P]^{c}_{l',r}}}$[NS-MOVE$_{na}$]~~~~$\mathsf{\frac{-}{T\rhd
0\xrightarrow{c?[(l:l'),r]}0}}$[NS-NULL$_{in2}$]}\\[3ex]

\normalsize{$\mathsf{\frac{T\rhd N_{1}\xrightarrow{c?[(l:l'),r]}
N'_{1}~T\rhd N_{2}\xrightarrow{c![(l:l'),r]} N'_{2}}{T\rhd
N_{1}|N_{2}\xrightarrow{c![(l:l'),r]}
N'_{1}|N'_{2}}}$[NS-MOVE]~~~~$\mathsf{\frac{T\rhd
N_{1}\xrightarrow{} N'_{1}}{T\rhd N_{1}|N_{2}\xrightarrow{}
N'_{1}|N_{2}}}$[NS-INT]}\\[2ex]
\hspace{-11ex}\scriptsize{\hspace{12ex}$\mathsf{T\rhd N_{2}|N_{1}\xrightarrow{c![(l:l'),r]} N'_{2}|N'_{1}}$\hspace{28ex}$\mathsf{T\rhd N_{2}|N_{1}\xrightarrow{} N_{2}|N'_{1}}$}\\[3ex]

$\mathsf{\frac{T\rhd N_{1}\xrightarrow{c?[(l:l'),r]} N'_{1}~~T\rhd
N_{2}\xrightarrow{c?[(l:l'),r]} N'_{2}}{T\rhd
N_{1}|N_{2}\xrightarrow{c?[(l:l'),r]}
N'_{1}|N'_{2}}}$[NS-MOVE$_{in}$]\\[4ex]
\footnotesize{\hspace{36ex}where $\mathsf{AIQ}$ is the set of processes of the form $\mathsf{(x)^{\delta}_{v}.P}$}\\
\bottomrule
\end{tabular}
\end{center}
\end{table}
Next rules NS-MOVE$_{ao}$, NS-MOVE$_{in1}$, NS-MOVE$_{in2}$,
NS-MOVE$_{in3}$, NS-MOVE$_{ai1}$, NS-MOVE$_{ai2}$, NS-MOVE$_{ai3}$
and NS-MOVE$_{na}$ are all used to deal with the node movement
events from different kinds of nodes. For example, rule
NS-MOVE$_{ao}$ depicts that an active transmitter located at
$\mathsf{l}$ moves to $\mathsf{l'}$ during its transmission over
channel $\mathsf{c}$ with radius $\mathsf{r}$. Then the behaviors of
surrounding nodes can be divided into three different cases
corresponding to NS-MOVE$_{in1}$, NS-MOVE$_{in2}$ and
NS-MOVE$_{in3}$ respectively. (1) For non-active nodes and active
receivers that are receiving over other channels or that are always
in or out of the transmission cell, they will remain unchanged. (2)
For active receivers that are originally within the transmission
cell, but later beyond it, they will receive an error. (3) For
active receivers that are in the reverse situation, they will get
interference. Analogously, rule NS-MOVE$_{ai1}$, NS-MOVE$_{ai3}$ and
NS-MOVE$_{ai2}$ have described the possible scenarios of an active
receiver that moves from $\mathsf{l}$ to $\mathsf{l'}$: (1) if the
active receiver moves out of the transmission cell, it will obtain
an error; (2) if the active receiver has always been within the
transmission cell and there is no more transmission in
$\mathsf{l'}$, it will continue to receive data normally; (3) if
there are more transmissions in $\mathsf{l'}$ apart from its
original one, the active receiver will get interference. Finally, we
can see from NS-MOVE$_{na}$ that for non-active nodes, they can move
arbitrarily without conditions and limitations.

Moreover, rule NS-PASS represents the responses of nodes as time
goes by. Rule NS-NULL$_{in1}$, NS-NULL$_{in2}$ and NS-NULL$_{pass}$
allow the empty network to receive data and evolve with time. At
last, the propagation of events through networks is portrayed by
rule NS-COM, NS-MOVE, NS-INT, NS-COM$_{in}$, NS-MOVE$_{in}$, and
NS-SYN. The first three denote that an event generated in a network
is propagated to the parallel network; while the later ones indicate
that two parallel networks receive the same event.
\vspace*{-2ex}
\section{Harmony Theorem}
\hspace*{4ex}The Harmony Theorem aims at proving that
the LTS-based semantics coincides with the RST-based semantics. With
this objective, the theorem has three parts. First, it shows that
the structural congruence respects the LTS, i.e., application of
structural congruence will not change the possible transitions. Then
it demonstrates that the RST behaves the same as the LTS, i.e., each
reduction in the RST has a corresponding transition in the LTS which
makes the resulting networks structurally congruent. In the end, it
testifies the converse also holds.

Before proving the theorem, there are some auxiliary lemmas that
portray the shape of processes able to perform a particular labelled
transition, and the shape of the derivative processes (see the Appendix).\\[2ex]
\textbf{Theorem 1} (\emph{Harmony Theorem}). Let $\mathsf{N}$ be a
network, and $\mathsf{T}$ a set of active transmitters.
\begin{enumerate}
 \item[(1)] If $\mathsf{T\rhd
N \xrightarrow{\mu} N'}$ and $\mathsf{N\equiv N_{1}}$, then there
exists $\mathsf{N'_{1}}$ such that $\mathsf{T\rhd N_{1}
\xrightarrow{\mu} N'_{1}\equiv N'}$.
 \item[(2)]
 \begin{enumerate}
   \item[(a)] If $\mathsf{T\rhd N\hookrightarrow^{c}_{l:l',r} N'}$, then $\mathsf{T\rhd N\xrightarrow{c![(l:l'),r]} N'_{1}\equiv N'}$.
   \item[(b)] If $\mathsf{T\rhd N\hookrightarrow^{c}_{l,r} N'}$, then
   there are $\mathsf{v}$ and $\mathsf{\delta}$ such that $\mathsf{T\rhd N\xrightarrow{c!v:\delta[l,r]} N'_{1}\equiv N'}$.
   \item[(c)] If $\mathsf{T\rhd N\hookrightarrow^{\sigma} N'}$, then $\mathsf{T\rhd N\xrightarrow{\sigma} N'_{1}\equiv N'}$.
   \item[(d)] If $\mathsf{T\rhd N\hookrightarrow N'}$, then $\mathsf{T\rhd N\xrightarrow{} N'_{1}\equiv N'}$.
 \end{enumerate}
 \item[(3)] For each item in (2), the reverse also holds.
\end{enumerate}
\begin{table}[!t]
\begin{center}
\begin{tabular}{>{\columncolor{mygray}\sf}c>{\columncolor{mygray}\sf}c>{\columncolor{mygray}\sf}c>{\columncolor{mygray}\sf}c}
\toprule
\rowcolor{white}\hspace*{-1.6ex}\textbf{Table 7.} Extended Syntax for Processes\\
\hline \hspace*{-25ex} $\mathsf{P \overset{def}=...}$ &
\hspace{-15ex} old processes & \hspace*{10.5ex}
$\mathsf{|\hspace{1ex}\lceil in(x).P\rceil^{t} Q}$ &
\hspace{5ex} input with timeout\\

\hspace*{-21.5ex} $\mathsf{|\hspace*{1ex}\sigma.P}$ & \hspace{-23ex}
delay & \hspace*{5.2ex} $\mathsf{|\hspace*{0.5ex}\vartriangleright
c.P}$ & \hspace{1ex}
channel switch\\

\hspace*{-16.5ex} $\mathsf{|\hspace{1ex}[e]Q_{1},Q_{2}}$ &
\hspace{-21.5ex} choice & \hspace*{5.2ex}
$\mathsf{|\hspace{1ex}H(\overrightarrow{u})}$ & \hspace{-4.5ex}
recursion\\
& \multicolumn{3}{>{\columncolor{mygray}\sf}c}{\hspace*{10ex}where $\mathsf{t}$ is a positive integer greater than 0}\\
\bottomrule
\end{tabular}
\end{center}
\end{table}
\textbf{Proof.} Now we prove the three points in
sequence.
\begin{enumerate}
 \item[(1)] The equivalence is defined in terms of commutativity, associativity, and identity over the empty network. First commutativity is guaranteed since rule NS-COM, NS-MOVE
 and NS-INT are symmetric, and rule NS-COM$_{in}$, NS-MOVE$_{in}$ and NS-SYS are self-symmetric. Identity over the empty network conserves
 since, owing to rule NS-NULL$_{in1}$, NS-NULL$_{in2}$ and
 NS-NULL$_{pass}$, the empty network can perform any labels of the
 form $\mathsf{\xrightarrow{c?v:\delta[l,r]}}$, $\mathsf{\xrightarrow{c?[(l:l'),r]}}$, and $\mathsf{\xrightarrow{\sigma}}$, which serve as neutral element of parallel
 composition. Finally as the operations for parallel composition are associative
 and the network structure is always preserved, therefore associativity is also ensured.
 \item[(2)]
 As proofs for the four statements are similar, we only take
 (a) as an example. \\
 (a) The proof is by rule induction on the derivation of
 $\mathsf{T\rhd N\hookrightarrow^{c}_{l:l',r} N'}$. \\
 \hspace*{4ex}First we consider rule RST-MOVE-AO, the proof for this case is by induction on the
 size of $\mathsf{I\cup J\cup K}$. The base case
 is $\mathsf{I\cup J\cup K=\emptyset}$, using rule NS-MOVE$_{ao}$.
 In the inductive case, we randomly choose an element
 $\mathsf{h}$ from $\mathsf{I\cup J\cup K}$. Remember that by the
 inductive hypothesis, we already have a transition with label
 $\mathsf{\xrightarrow{c![(l:l'),r]}}$. Below are different
 cases according to which set $\mathsf{h}$ belongs to. Suppose
 $\mathsf{h\in I}$, then we can apply rule NS-MOVE$_{in1}$
 since $\mathsf{d(l,l_{i})\leq r\wedge d(l',l_{i})\leq r}$
 from the premise of rule RST-MOVE-AO. Thus the desired
 transition can be proved using rule NS-MOVE.
 Suppose now $\mathsf{h\in J}$, we can use rule PS-IN$_{err}$
 to derive $\mathsf{(x_{j})^{\delta}_{v}.P_{j}\xrightarrow{?\epsilon}
 P_{j}\{\epsilon/x_{j}\}}$, and then rule NS-MOVE$_{in2}$ to
 derive a transition with label $\mathsf{\xrightarrow{c?[(l:l'),r]}}$.
 Hence the desired format can be arrived using rule NS-MOVE.
 Finally suppose $\mathsf{h\in K}$, we can use rule
 PS-IN$_{interfere}$ to derive $\mathsf{(x_{k})^{\delta_{k}}_{v_{k}}.P_{k}\xrightarrow{?\bot}
 P_{k}\{\bot/x_{k}\}}$. This transition can be lifted up to the
 network level using rule NS-MOVE$_{in3}$ because
 $\mathsf{d(l,l_{k})>r\wedge d(l',l_{k})\leq r}$ from the precondition of rule RST-MOVE-AO.
 Similarly, using rule NS-MOVE, we can get the desired transition.\\
 \hspace*{4ex}The proof is analogous for rule RST-CONT-MOVE, since $\mathsf{(l,r,c)\NotDownarrowTwo_{ai} N''\wedge}$\\
 $\mathsf{(l',r,c)\NotDownarrowTwo_{ai} N''}$ ensures that all the active input nodes
 satisfy the conditions of NS-MOVE$_{in1}$. As for non-active
 input nodes, rule NS-MOVE$_{in1}$ can also be applied.\\
 \hspace*{4ex}Rule RST-CONGR can be simulated by the first part of
 the theorem.
 \item[(3)]
 The proof for (a) and (b) are based on Lemma 2 and Lemma 4
 respectively. The proof for (d) is straightforward. Now we consider
 the proof for (c).\\
 The proof is based on Lemma 5. If $\mathsf{I\cup J\cup
 K=\emptyset}$, the desired reduction can be derived using rule
 RST-PASS-NULL and RST-CONGR. Otherwise, for each element
 $\mathsf{i}$ in $\mathsf{I}$ and for each element $\mathsf{k}$ in
 $\mathsf{K}$, first apply rule RST-SENDING and RST-PASS-NA
 respectively, then employ rule RST-CONT-PASS and RST-CONGR to get
 the desired reduction. \hfill $\Box$
\end{enumerate}
\vspace*{-2ex}
\section{The Extended Language}
\hspace*{4ex}So far we have considered the subset of TCMN with only
the operators that are necessary for communication. Now we present
some extensions: a series of processes are added in Table 7, while
the syntax for others remains the same.

First of all, the input construct is replaced by the \emph{input
with timeout} construct in $\mathsf{\lceil in(x).P\rceil^{t} Q}$.
This process is waiting for receiving a value, if the value arrives
before the end of the $\mathsf{t}$ time units, the process evolves
into an active receiver; otherwise, the process continues as
$\mathsf{Q}$. Process $\mathsf{\sigma.P}$ stands for sleeping for
one time unit while $\mathsf{\vartriangleright c.P}$ represents a
process that decides to switch its communication channel to
$\mathsf{c}$, and then continues as $\mathsf{P}$. The construct
$\mathsf{[e]Q_{1},Q_{2}}$ behaves as $\mathsf{Q_{1}}$ if
$\mathsf{e=true}$ and as $\mathsf{Q_{2}}$ otherwise. Here,
$\mathsf{e}$ is a boolean value expression. Finally,
$\mathsf{H(\overrightarrow{u})}$ denotes a process defined via a
(possibly recursive) definition
$\mathsf{H(\overrightarrow{x})\overset{def}=Q}$, with
$\mathsf{|\overrightarrow{x}|=|\overrightarrow{u}|}$, where
$\mathsf{\overrightarrow{x}}$ contains all free variables of
$\mathsf{Q}$.

We only provide the addition of the new operators to the RST
semantics, since this is the simpler one and the one that we will
use in Section 7. However the operators can be introduced in a
similar way into the LTS semantics.

Before updating the RST semantics, a new structural congruence rule
is appended: \\
\hspace*{8ex}$\mathsf{n[H(\overrightarrow{u})]^{c}_{l,r}\equiv
n[Q\{\overrightarrow{u}/\overrightarrow{x}\}]^{c}_{l,r}}$ if
$\mathsf{H(\overrightarrow{x})\overset{def}=Q\wedge
|\overrightarrow{x}|=|\overrightarrow{u}|}$

\begin{table}[!t]
\begin{center}
\begin{tabular}{>{\columncolor{mygray}\sf}c}
\toprule
\rowcolor{white}\hspace*{-42ex} \textbf{Table 8.} Extended Reduction Semantics\\
\hline\\[-0.5ex]
[RST-BEGIN] \\[0.5em]
\normalsize{$\mathsf{\frac{\forall h\in I\cup J\cup K.d(l,l_{h})\leq
r~~\forall i\in I.T|_{l_{i},c}=\emptyset~~\forall j \in
J.T|_{l_{j},c}\neq \emptyset}{T\rhd n[out\langle
u\rangle.P]^{c}_{l,r}|\prod\limits_{h\in I\cup J}n_{h}[\lceil
in(x_{h}).P_{h}\rceil^{t_{h}}
Q_{h}]^{c}_{l_{h},r_{h}}|\prod\limits_{k\in
K}n_{k}[(x_{k})^{\delta_{k}}_{v_{k}}.P_{k}]^{c}_{l_{k},r_{k}}\hookrightarrow
^{c}_{l,r}}}$}\\[1.5em]
\scriptsize{\hspace*{0ex}$\mathsf{n[\langle \ThreeQuotersLeft u
\ThreeQuotersRight \rangle^{\Lbag
u\Rbag}.P]^{c}_{l,r}|\prod\limits_{i\in I}n_{i}[(x_{i})^{\Lbag
u\Rbag}_{\ThreeQuotersLeft u
\ThreeQuotersRight}.P_{i}]^{c}_{l_{i},r_{i}}|\prod\limits_{j\in
J}n_{j}[\lceil in(x_{j}).P_{j}\rceil^{t_{j}}
Q_{j}]^{c}_{l_{j},r_{j}}|\prod\limits_{k\in
K}n_{k}[P_{k}\{\bot/x_{k}\}]^{c}_{l_{k},r_{k}}}$}\\[2em]

\hspace{6ex}[RST-INPUT-DELAY]\hspace{10ex}[RST-TIMEOUT]\hspace{4ex}[RST-PASS-DELAY]\\[0.5em]
\normalsize{$\mathsf{\frac{t>1}{T\rhd n[\lceil in(x).P\rceil^{t}
Q]^{c}_{l,r}\hookrightarrow^{\sigma} n[\lceil in(x).P\rceil^{t-1}
Q]^{c}_{l,r}}}$~~~~$\mathsf{\frac{t=1}{T\rhd n[\lceil
in(x).P\rceil^{t} Q]^{c}_{l,r}\hookrightarrow^{\sigma}
n[Q]^{c}_{l,r}}}$}~~~~\scriptsize{$\mathsf{T\rhd n[\sigma.P]^{c}_{l,r}\hookrightarrow^{\sigma} n[P]^{c}_{l,r}}$}\\[4ex]

\hspace{-4ex}[RST-PASS-NA]\hspace{2ex}[RST-SWITCH]\hspace{4ex}[RST-IF-TRUE]\hspace{5ex}[RST-IF-FALSE]\\[0.5em]
\normalsize{$\mathsf{\frac{P\notin DIQ}{T\rhd n[P]^{c}_{l,r}\hookrightarrow^{\sigma} n[P]^{c}_{l,r}}}$}~~\scriptsize{$\mathsf{T\rhd n[\vartriangleright c'.P]^{c}_{l,r}\hookrightarrow n[P]^{c'}_{l,r}}$}~~\normalsize{$\mathsf{\frac{e=true}{T\rhd n[[e]Q_{1},Q_{2}]^{c}_{l,r}\hookrightarrow n[Q_{1}]^{c}_{l,r}}}$~~$\mathsf{\frac{e=false}{T\rhd n[[e]Q_{1},Q_{2}]^{c}_{l,r}\hookrightarrow n[Q_{2}]^{c}_{l,r}}}$}\\[3ex]
\footnotesize{\hspace{30ex}where $\mathsf{DIQ}$ is the set of processes of the form $\mathsf{\sigma.P}$ or $\mathsf{\lceil in(x).P\rceil^{t} Q}$}\\
\bottomrule
\end{tabular}
\end{center}
\end{table}
The additional reduction rules are shown in Table 8. Rule RST-BEGIN
is as before, except that input is substituted by input with
timeout. In RST-TIMEOUT, a timeout fires if no reception has started
before the end of the current instant of time. For processes of the
form $\mathsf{\lceil in(x).P\rceil^{t} Q}$ and $\mathsf{\sigma.P}$,
rule RST-INPUT-DELAY and RST-PASS-DELAY model the sleeping for one
time unit respectively. Rule RST-PASS-NA is a modification of the
former one: non-active processes other than those of the form
$\mathsf{\lceil in(x).P\rceil^{t} Q}$ and $\mathsf{\sigma.P}$ have
no reaction to the time passing event. Then the remaining rules are
self-explanatory.
\vspace*{-1ex}
\section{Case Study}
\hspace*{4ex}We start this section by taking some MAC-layer
protocols: CSMA and MACA/R-T as examples to show the expressiveness
of our calculus.
\vspace*{-1ex}
\subsection{Carrier Sense Multiple Access}
The \emph{Carrier Sense Multiple Access} (CSMA) scheme is a widely
used MAC-layer protocol. In this protocol, each device senses the
channel (physical carrier) before its transmission. If the channel
is free, the sender starts the transmission immediately; otherwise
the device keeps monitoring the channel until it becomes idle and
then starts the transmission.

We can easily model the carrier sense action of
CSMA scheme by the process defined below:\\[2ex]
\hspace*{8ex}$\mathsf{Send(l,c,u)\overset{def}=[T|_{l,c}=\emptyset]out\langle
u\rangle,\sigma.Send(l,c,u)}$\\[2ex]
Let us represent some reduction traces for a network where nodes
adopt the CSMA protocol. These traces indicate that the CSMA
protocol does not address the issue of node mobility. When an active
transmitter moves to the reception range of an occupied receiver,
any transmission of the intruding node may cause interference with
the ongoing transmission. Similarly, if an active receiver moves in
the transmission cell of another transmitter, the transmission of
the new transmitter will also interfere with the original one.
Further, interference may as well occur when different packages are
targeted at the same receiver simultaneously.\\[2ex]
\begin{figure*}[htb]
\centering
\includegraphics[width=0.23\textwidth]{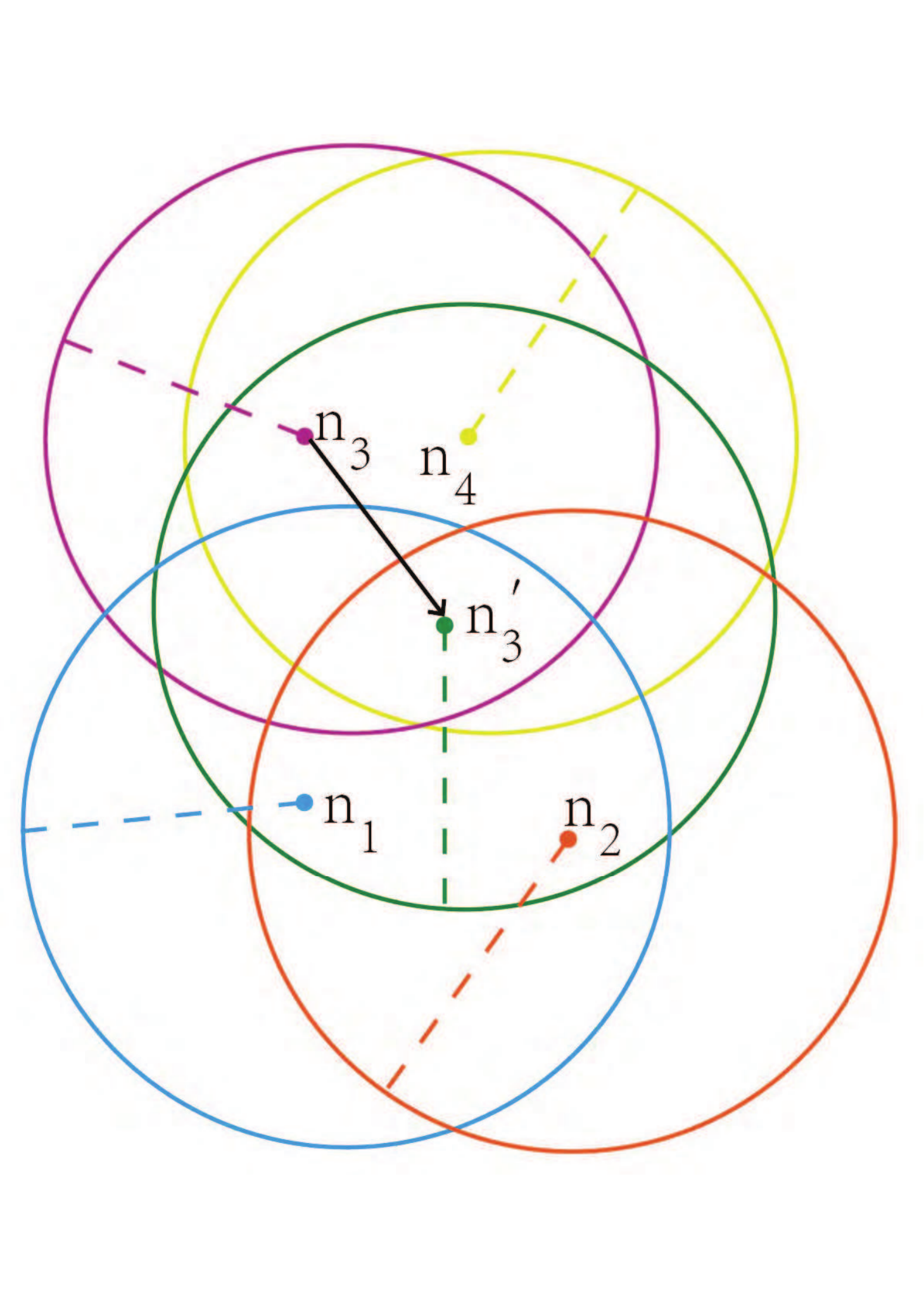}
\caption{Network topology of Lemma 1-2 and Example 1-3}
\label{fig:ropart}
\end{figure*}
\textbf{Example 1} (\emph{Interference}). This example represents an
active transmitter $\mathsf{n_{3}}$ moves to $\mathsf{n^{'}_{3}}$
during its communication with node $\mathsf{n_{4}}$. Due to this
event, the active receiver $\mathsf{n_{2}}$ which is receiving data
from node $\mathsf{n_{1}}$ gets an interference since it passively
enters the
transmission cell of $\mathsf{n^{'}_{3}}$. The network is: \\[2ex]
\hspace*{4ex}$\mathsf{N\overset{def}=n_{1}[\langle v_{1}
\rangle^{\delta_{1}}]^{c}_{l_{1},r_{1}}~|~~n_{2}[(x_{2})^{\delta_{1}}_{v_{1}}.P_{1}]^{c}_{l_{2},r_{2}}~|~n_{3}[\langle
v_{3}
\rangle^{\delta_{3}}]^{c}_{l_{3},r_{3}}~|~n_{4}[(x_{4})^{\delta_{3}}_{v_{3}}.P_{4}]^{c}_{l_{4},r_{4}}}$\\[2ex]
where $\mathsf{n_{2}}$ and $\mathsf{n^{'}_{3}}$ are in the
transmission cell of $\mathsf{n_{1}}$, just as $\mathsf{n_{4}}$ in
$\mathsf{n_{3}}$, $\mathsf{n_{1}}$ and $\mathsf{n_{2}}$ together
with $\mathsf{n_{4}}$ in $\mathsf{n^{'}_{3}}$ (as shown in Fig.1).\\[0.8ex]
We present a possible reduction trace, and it is easily understood.\\[0.8ex]
\footnotesize{\hspace*{2ex}$\mathsf{\{(l_{1},r_{1},c),(l_{3},r_{3},c)\}\rhd
N\hookrightarrow^{c}_{l_{3}:l'_{3},r_{3}} n_{1}[\langle v_{1}
\rangle^{\delta_{1}}]^{c}_{l_{1},r_{1}}
|n_{2}[P_{2}\{\bot/x_{2}\}]^{c}_{l_{2},r_{2}}|n_{3}[\langle v_{3}
\rangle^{\delta_{3}}]^{c}_{l'_{3},r_{3}}|n_{4}[(x_{4})^{\delta_{3}}_{v_{3}}.P_{4}]}$}\\[0.8ex]
\normalsize Analogously, if $\mathsf{n_{3}}$ is an active receiver
and moves to $\mathsf{n^{'}_{3}}$ during its reception from
$\mathsf{n_{4}}$.
Then the transmission of $\mathsf{n_{1}}$ to $\mathsf{n_{2}}$ will also interfere with $\mathsf{n_{4}}$ to $\mathsf{n^{'}_{3}}$.\hfill $\Box$\\[1ex]
\textbf{Example 2} (\emph{Interference}). This example indicates
interference caused by the simultaneous transmission of two
different packages. Let us consider now a different network:\\[2ex]
\hspace*{4ex}$\mathsf{N\overset{def}=n_{1}[Send(l_{1},c,u_{1})]^{c}_{l_{1},r_{1}}~|~n_{2}[Send(l_{2},c,u_{2})]^{c}_{l_{2},r_{2}}~|~n_{3}[in(x).P]^{c}_{l'_{3},r_{3}}}$\\[2ex]
where $\mathsf{n_{2}}$ and $\mathsf{n^{'}_{3}}$ are in the
transmission cell of $\mathsf{n_{1}}$, just as $\mathsf{n_{1}}$ and
$\mathsf{n^{'}_{3}}$ in $\mathsf{n_{2}}$ (see Fig.1).\\[0.8ex]
A possible reduction trace is given:\\[0.8ex]
\footnotesize{\hspace*{2ex}$\mathsf{\emptyset\rhd N\hookrightarrow
n_{1}[out\langle u_{1} \rangle]^{c}_{l_{1},r_{1}}
|n_{2}[Send(l_{2},c,u_{2})]^{c}_{l_{2},r_{2}}|n_{3}[in(x).P]^{c}_{l'_{3},r_{3}}\overset{def}=N_{1}}$\\[1ex]
\hspace*{2ex}$\mathsf{\emptyset\rhd N_{1}\hookrightarrow
n_{1}[out\langle u_{1} \rangle]^{c}_{l_{1},r_{1}}
|n_{2}[out\langle u_{2} \rangle]^{c}_{l_{2},r_{2}}|n_{3}[in(x).P]^{c}_{l'_{3},r_{3}}\overset{def}=N_{2}}$\\[1ex]
Assign $\mathsf{v_{1}=\ThreeQuotersLeft u_{1} \ThreeQuotersRight}$ and $\mathsf{\delta_{1}=\Lbag u_{1}\Rbag}$, then\\[1ex]
\hspace*{2ex}$\mathsf{\emptyset\rhd
N_{2}\hookrightarrow^{c}_{l_{1},r_{1}} n_{1}[\langle v_{1}
\rangle^{\delta_{1}}]^{c}_{l_{1},r_{1}}
|n_{2}[out\langle u_{2} \rangle]^{c}_{l_{2},r_{2}}|n_{3}[(x)^{\delta_{1}}_{v_{1}}.P]^{c}_{l'_{3},r_{3}}\overset{def}=N_{3}}$\\[1ex]
Assign $\mathsf{v_{2}=\ThreeQuotersLeft u_{2} \ThreeQuotersRight}$ and $\mathsf{\delta_{2}=\Lbag u_{2}\Rbag}$, then\\[1ex]
\hspace*{2ex}$\mathsf{\{(l_{1},r_{1},c)\}\rhd
N_{3}\hookrightarrow^{c}_{l_{2},r_{2}} n_{1}[\langle v_{1}
\rangle^{\delta_{1}}]^{c}_{l_{1},r_{1}} |n_{2}[\langle v_{2}
\rangle^{\delta_{2}}]^{c}_{l_{2},r_{2}}|n_{3}[P\{\bot/x\}]^{c}_{l'_{3},r_{3}}}$}\\[1ex]
\normalsize Here $\mathsf{n_{1}}$ senses the channel free, and then
almost at the same time, $\mathsf{n_{2}}$ also finds the channel
available, so they begin to transmit data successively. Therefore an
interference is generated at $\mathsf{n^{'}_{3}}$.\hfill$\Box$
\vspace*{-1ex}
\subsection{MACA/R-T}
\begin{table}[!t]
\begin{center}
\begin{tabular}{>{\columncolor{mygray}\sf}c}
\toprule
\rowcolor{white}\hspace*{-54ex}\textbf{Table 9.} MACA/R-T\\
\hline
\hspace{-8.5ex}$\mathsf{SND(sid,rid,u)\overset{def}=\vartriangleright
c_{r,rid}.out\langle
\{sid,rid,rts,\Lbag\{sid,rid,end,u\}\Rbag\}\rangle.}$ \\
\hspace{-19ex}$\mathsf{\vartriangleright c_{s,rid}.\lceil in(x).}$ \\
\hspace{26ex}$\mathsf{[fst(x)=rid\wedge snd(x)=sid\wedge trd(x)=cts]}$ \\
\hspace{24.5ex}$\mathsf{\vartriangleright c_{s,sid}.out\langle\{sid,rid,end,u\}\rangle.\vartriangleright c_{r,sid},}$ \\
\hspace{5.3ex}$\mathsf{SND(sid,rid,u)\rceil^{t}}$ \\
\hspace{-3.5ex}$\mathsf{SND(sid,rid,u)}$ \\

\hspace{-49ex}$\mathsf{RCV(id,q)\overset{def}=\lceil in(x).}$ \\
\hspace{-13ex}$\mathsf{[snd(x)=id\wedge trd(x)=rts]}$ \\
\hspace{-3ex}$\mathsf{\vartriangleright c_{s,id}.out\langle
\{id,fst(x),cts,fth(x)\}\rangle.}$ \\
\hspace{-18.5ex}$\mathsf{\vartriangleright c_{s,fst(x)}.\lceil in(y).}$ \\
\hspace{31ex}$\mathsf{[snd(y)=id\wedge fst(y)=fst(x)\wedge trd(y)=end]}$ \\
\hspace{15ex}$\mathsf{RCV(id,push(q,fth(y))),}$ \\
\hspace{10ex}$\mathsf{\vartriangleright c_{r,id}.RCV(id,q)\rceil^{t}}$ \\
\hspace{2ex}$\mathsf{\vartriangleright c_{r,id}.RCV(id,q),}$ \\
\hspace{-24.8ex}$\mathsf{RCV(id,q)\rceil^{t}}$ \\
\hspace{-32ex}$\mathsf{RCV(id,q)}$ \\
\bottomrule
\end{tabular}
\end{center}
\end{table}
The \emph{Receiver-Transmitter-Based Multiple Access with Collision
Avoidance Protocol} (MACA/R-T) is a promising protocol used in
MANETs. In MACA/R-T, all mobile nodes in the network agree to a set
of pre-specified channels, e.g., a node $\mathsf{id}$ is assigned
with $\mathsf{c_{r,id}}$ and $\mathsf{c_{s,id}}$ as its receiver and
transmitter channels respectively. At the idle stage, all nodes will
tune their receivers to their own receiver channel. When node
$\mathsf{sid}$ wants to send a data package to node $\mathsf{rid}$,
node $\mathsf{sid}$ first sends a short control packet RST
(request-to-send, which includes the sender id, the receiver id and
the transmission duration of the data package) to node
$\mathsf{rid}$ over channel $\mathsf{c_{r,rid}}$ and then tunes its
receiver to channel $\mathsf{c_{s,rid}}$ to wait for a control
packet CTS (clear-to-send, which includes the same duration
information) from node $\mathsf{rid}$. Upon receiving the RTS, node
$\mathsf{rid}$ will send a CTS over channel $\mathsf{c_{s,rid}}$ and
tune its receiver to channel $\mathsf{c_{s,sid}}$ for the data
package. Finally, node $\mathsf{sid}$ receives the CTS and sends the
data package to node $\mathsf{rid}$ over channel
$\mathsf{c_{s,sid}}$.

In Table 9, we provide an encoding of a sender and a receiver
process in our TCMN with respect to the MACA/R-T protocol. We assume
that the receiver has a queue to store the received packages, with
an operation $\mathsf{push}$ to insert an element. We also use
four-tuples as values, with constructor $\mathsf{\{\_,\_,\_,\_\}}$
and destructors $\mathsf{fst}$, $\mathsf{snd}$, $\mathsf{trd}$ and
$\mathsf{fth}$, retrieving the first, second, third and fourth
component separately. We indicate with $\mathsf{t}$ the maximum time
for a data package from the sender to arrive at the receiver.

The sender process $\mathsf{SND(sid,rid,u)}$ runs at node
$\mathsf{sid}$ and intends to transmit the value $\mathsf{u}$ to
node $\mathsf{rid}$. The process first switches its transmitter
channel to $\mathsf{c_{r,rid}}$ and then sends a RTS packet. After
that, it waits for the CTS packet. If the CTS packet is not received
before the end of the $\mathsf{t}$ time units, the process will move
to itself and restart the transmission. On the other hand, if the
CTS packet is received before the timeout, the data package
$\mathsf{\{sid,rid,end,u\}}$ is transmitted over channel
$\mathsf{c_{s,sid}}$ and the sender finishes the transmission.

The receiver process $\mathsf{RCV(id,q)}$ is supposed to run at node
$\mathsf{id}$ waiting for a RTS packet. If the RTS packet, with
destination $\mathsf{id}$, arrives before the timeout, the receiver
switches its transmitter channel to $\mathsf{c_{s,id}}$ and then
replies with a CTS packet as well as waits for the data package over
channel $\mathsf{c_{s,fst(x)}}$. Otherwise, the receiver aborts the
current reception and resets to process $\mathsf{RCV(id,q)}$.

We show below that the MACA/R-T protocol is robust against node
mobility, i.e., node movement will not give rise to communication
interference. When an active transmitter moves to the reception
range of an occupied receiver, the transmission of the intruding
node will not interfere with the ongoing one. Besides, if an active
receiver moves in the transmission cell of another transmitter, the
transmission of the new transmitter will not interfere with the
original one.\\[2ex]
\textbf{Lemma 1} Suppose when node $\mathsf{n_{1}}$ is transmitting
to node $\mathsf{n_{2}}$ and node $\mathsf{n_{3}}$ is transmitting
to node $\mathsf{n_{4}}$, $\mathsf{n_{3}}$ moves to
$\mathsf{n^{'}_{3}}$. The network topology is shown in Fig.1,
$\mathsf{n_{2}}$ and $\mathsf{n^{'}_{3}}$ are in the transmission
cell of $\mathsf{n_{1}}$, just as $\mathsf{n_{4}}$ in
$\mathsf{n_{3}}$, $\mathsf{n_{1}}$ and $\mathsf{n_{2}}$ together
with $\mathsf{n_{4}}$ in $\mathsf{n^{'}_{3}}$. Then the transmission
of $\mathsf{n^{'}_{3}}$ to $\mathsf{n_{4}}$ will not interfere with
that of $\mathsf{n_{1}}$ to $\mathsf{n_{2}}$. \\[0.8ex]
\textbf{Proof.} Remember that only when an active receiver has
received more than one transmission over the same channel, does the
receiver get interference.

There are three kinds of packages in the MACA/R-T protocol: RTS, CTS
and data, which are transmitted over channels $\mathsf{c_{r,rid}}$,
$\mathsf{c_{s,rid}}$, and $\mathsf{c_{s,sid}}$ respectively.
According to the package types that $\mathsf{n_{1}}$ and
$\mathsf{n_{3}}$ are sending, all the possible cases of the active
transmitter $\mathsf{n_{3}}$ moves to $\mathsf{n^{'}_{3}}$ are
listed in the table below.
\begin{table}[h]
\begin{center} {\footnotesize
\begin{tabular}{|c|c|c|c|c|}
\hline $\mathsf{n_{3}}\dashrightarrow \mathsf{n_{4}}$ &
$\mathsf{n^{'}_{3}}\dashrightarrow \mathsf{n_{4}}$ &
$\mathsf{n_{1}}\dashrightarrow \mathsf{n_{2}}$ &
Interference at $\mathsf{n_{2}}$ & Reasons \\
\hline $\mathsf{c_{r,n_{4}}.RTS}$ & $\mathsf{c_{r,n_{4}}.RTS}$ &
$\mathsf{c_{r,n_{2}}.RTS}$  &  No & Different channels \\
$\mathsf{c_{r,n_{4}}.RTS}$ & $\mathsf{c_{r,n_{4}}.RTS}$ &
$\mathsf{c_{s,n_{2}}.CTS}$  &  No & Different channels \\
$\mathsf{c_{r,n_{4}}.RTS}$ & $\mathsf{c_{r,n_{4}}.RTS}$ &
$\mathsf{c_{s,n_{1}}.data}$  &  No & Different channels \\
$\mathsf{c_{s,n_{4}}.CTS}$ & $\mathsf{c_{s,n_{4}}.CTS}$ &
$\mathsf{c_{r,n_{2}}.RTS}$  &  No & Different channels \\
$\mathsf{c_{s,n_{4}}.CTS}$ & $\mathsf{c_{s,n_{4}}.CTS}$ &
$\mathsf{c_{s,n_{2}}.CTS}$  &  No & Different channels \\
$\mathsf{c_{s,n_{4}}.CTS}$ & $\mathsf{c_{s,n_{4}}.CTS}$ &
$\mathsf{c_{s,n_{1}}.data}$  &  No & Different channels \\
$\mathsf{c_{s,n_{3}}.data}$ & $\mathsf{c_{s,n_{3}}.data}$ &
$\mathsf{c_{r,n_{2}}.RTS}$  &  No & Different channels \\
$\mathsf{c_{s,n_{3}}.data}$ & $\mathsf{c_{s,n_{3}}.data}$ &
$\mathsf{c_{s,n_{2}}.CTS}$  &  No & Different channels \\
$\mathsf{c_{s,n_{3}}.data}$ & $\mathsf{c_{s,n_{3}}.data}$ &
$\mathsf{c_{s,n_{1}}.data}$  &  No & Different channels \\
\hline
\end{tabular} }
\end{center}
\end{table}
\vspace*{-1ex} \hspace*{2ex}
We can see that when an active
transmitter (e.g., $\mathsf{n_{3}}$) moves to the reception range of
an occupied receiver (e.g., $\mathsf{n_{2}}$), due to the different
tranmission channels, the transmission of the intruding node will
not interfere with the
ongoing one.\hfill $\Box$ \\[2ex]
\textbf{Lemma 2} Suppose when node $\mathsf{n_{1}}$ is transmitting
to node $\mathsf{n_{2}}$ and node $\mathsf{n_{4}}$ is transmitting
to node $\mathsf{n_{3}}$, $\mathsf{n_{3}}$ moves to
$\mathsf{n^{'}_{3}}$. As shown in Fig.1, $\mathsf{n_{2}}$ and
$\mathsf{n^{'}_{3}}$ are in the transmission cell of
$\mathsf{n_{1}}$, just like $\mathsf{n_{3}}$ and
$\mathsf{n^{'}_{3}}$ in $\mathsf{n_{4}}$. Then the transmission of
$\mathsf{n_{1}}$ to $\mathsf{n_{2}}$ will not interfere with
that of $\mathsf{n_{4}}$ to $\mathsf{n^{'}_{3}}$. \\[0.8ex]
\textbf{Proof.} The proof is similar to the one for Lemma 1, and we
conclude that if an active receiver (e.g., $\mathsf{n_{3}}$) moves
in the transmission cell of another transmitter (e.g.,
$\mathsf{n_{1}}$), the transmission of the new transmitter will not
interfere with the original one.\hfill $\Box$ \\[2ex]
Nevertheless, in the MACA/R-T protocol, interference may still occur
when different RTS packets are targeted at the same receiver
simultaneously.\\[2ex]
\textbf{Example 3} (\emph{Interference}). Let's consider the
network:\\[2ex]
\hspace*{2ex}$\mathsf{N\overset{def}=n_{1}[SND(n_{1},n_{3},u_{1})]^{c_{r,n_{1}}}_{l_{1},r_{1}}~|~n_{2}[SND(n_{2},n_{3},u_{2})]^{c_{r,n_{2}}}_{l_{2},r_{2}}~|~n_{3}[RCV(n_{3},[~])]^{c_{r,n_{3}}}_{l^{'}_{3},r_{3}}}$\\[1.5ex]
where $\mathsf{n_{2}}$ and $\mathsf{n^{'}_{3}}$ are in the
transmission cell of $\mathsf{n_{1}}$, as $\mathsf{n_{1}}$ and
$\mathsf{n^{'}_{3}}$ in $\mathsf{n_{2}}$ (see Fig.1).\\
Here we present a possible reduction trace:\\[0.8ex]
\footnotesize{\hspace*{2ex}$\mathsf{\emptyset \rhd N\hookrightarrow\hookrightarrow\hookrightarrow^{c_{r,n_{3}}}_{l_{1},r_{1}}\hookrightarrow^{c_{r,n_{3}}}_{l_{2},r_{2}}}$}\\[0.8ex]
\normalsize Initially, $\mathsf{n_{1}}$ tunes its receiver to
channel $\mathsf{c_{r,n_{3}}}$ and sends a RTS packet to
$\mathsf{n^{'}_{3}}$. Almost at the same time, $\mathsf{n_{2}}$ also
tunes its receiver to channel $\mathsf{c_{r,n_{3}}}$ and sends a RTS
packet to $\mathsf{n^{'}_{3}}$ which unfortunately results in an
interference at $\mathsf{n^{'}_{3}}$.\hfill $\Box$
\vspace*{-1ex}
\section{Conclusions and Future Work}
 \hspace*{4ex}\normalsize{In this paper, we have proposed a timed calculus for
 mobile ad hoc networks paying particular attention to local
 broadcast, node mobility and communication interference. Then
 the operational semantics of our calculus is given both in
 terms of a Reduction Semantics and in terms of a Labelled
 Transition Systems. We have also proved that these two semantics
 coincide. Finally, we extend our core language by adding
 some new operators to model the CSMA and MACA/R-T protocol.
 And we have demonstrated that the former doesn't address the issue of node
 mobility while the latter is robust against node mobility.}

 In the future, a quantity of developments are possible. First,
 we would try to establish adequate Behavioral Equivalences which
 define when two terms have the same observable behavior.
 One possible approach is via UTP method, so as to investigate the
 denotational semantics for mobile ad hoc networks. Second,
 we would also like to study a set of algebraic
 laws, which can represent the features of mobile ad hoc networks.
\providecommand{\urlalt}[2]{\href{#1}{#2}}
\providecommand{\doi}[1]{doi:\urlalt{http://dx.doi.org/#1}{#1}}
\bibliographystyle{eptcs}

\vspace*{-2ex}
\appendix
\section{Appendix}
\textbf{Lemma 1.} If $\mathsf{T\rhd N\xrightarrow{c?[(l:l'),r]}N'}$, then\\[1.5ex]
\hspace*{4ex}$\mathsf{N\equiv \prod\limits_{i\in
I}n_{i}[(x_{i})^{\delta}_{v}.P_{i}]^{c}_{l_{i},r_{i}}|\prod\limits_{j\in
J}n_{j}[(x_{j})^{\delta}_{v}.P_{j}]^{c}_{l_{j},r_{j}}|\prod\limits_{k\in
K}n_{k}[(x_{k})^{\delta_{k}}_{v_{k}}.P_{k}]^{c}_{l_{k},r_{k}}|N''}$\\[1ex]
where $\mathsf{\forall i\in I.d(l,l_{i})\leq r\wedge d(l',l_{i})\leq
r},~~\mathsf{\forall j\in J.d(l,l_{j})\leq r\wedge d(l',l_{j})>r},~~\mathsf{\forall k\in K.d(l,l_{k})>r\wedge d(l',l_{k})\leq r}$\\[1.5ex]
and $\mathsf{(l,r,c)\NotDownarrowTwo_{ai}
N''\wedge (l',r,c)\NotDownarrowTwo_{ai} N''}$. Furthermore\\[1.2ex]
\hspace*{4ex}$\mathsf{N'\equiv \prod\limits_{i\in
I}n_{i}[(x_{i})^{\delta}_{v}.P_{i}]^{c}_{l_{i},r_{i}}|\prod\limits_{j\in
J}n_{j}[P_{j}\{\epsilon/x_{j}\}]^{c}_{l_{j},r_{j}}|\prod\limits_{k\in
K}n_{k}[P_{k}\{\bot/x_{k}\}]^{c}_{l_{k},r_{k}}|N''}$\\[1ex]
\textbf{Proof.} The proof is by rule induction on the derivation of
$\mathsf{T\rhd N\xrightarrow{c?[(l:l'),r]}N'}$. \\[1.2ex]
\emph{Rule NS-MOVE$_{in1}$} From the premise of the rule, we know
that either $\mathsf{n[Q]^{c}_{l,r}}$ is a non-active input node or
an active input node that always within or beyond the mobile
transmitter's transmission range. In the first and third case, the
corresponding node can be inserted into $\mathsf{N''}$, while in the
second case, $\mathsf{Q=Q'=(x_{i})^{\delta}_{v}.P_{i}}$, thus it
follows that
$\mathsf{I=\{i\},J=K=\emptyset}$, and $\mathsf{N''=0}$.\\[0.8ex]
\emph{Rule NS-MOVE$_{in2}$} Here too we know that $\mathsf{Q}$ is an
active input process and $\mathsf{Q\xrightarrow{?\epsilon}Q'}$, thus
by inspection on the LTS for processes, we get one case, rule
PS-IN$_{err}$. It corresponds to index $\mathsf{Q}$ with
$\mathsf{j\in J}$.\\[0.8ex]
\emph{Rule NS-MOVE$_{in3}$} Similarly, when $\mathsf{Q}$ is an
active input process, then $\mathsf{Q\xrightarrow{?\bot}Q'}$ can
only be derived using rule RS-IN$_{interfere}$. Hence, $\mathsf{Q}$
is indexed with $\mathsf{k\in K}$.\\[0.8ex]
\emph{Rule NS-MOVE$_{in}$} This is the inductive case. It brings the
corresponding sets of indices and the non-index part of the network
in the previous premises to the desired form. \hfill $\Box$ \\[2ex]
\textbf{Lemma 2.} If $\mathsf{T\rhd N\xrightarrow{c![(l:l'),r]}N'}$, then\\[1.5ex]
\hspace*{4ex}$\mathsf{N\equiv n[\langle v
\rangle^{\delta}.P]^{c}_{l,r}|\prod\limits_{i\in
I}n_{i}[(x_{i})^{\delta}_{v}.P_{i}]^{c}_{l_{i},r_{i}}|\prod\limits_{j\in
J}n_{j}[(x_{j})^{\delta}_{v}.P_{j}]^{c}_{l_{j},r_{j}}|\prod\limits_{k\in
K}n_{k}[(x_{k})^{\delta_{k}}_{v_{k}}.P_{k}]^{c}_{l_{k},r_{k}}|N''}$\\[1ex]
where $\mathsf{\forall i\in I.d(l,l_{i})\leq r\wedge d(l',l_{i})\leq
r},~~\mathsf{\forall j\in J.d(l,l_{j})\leq r\wedge d(l',l_{j})>r},~~
\mathsf{\forall k\in K.d(l,l_{k})>r\wedge d(l',l_{k})\leq r}$\\[1.5ex]
and $\mathsf{(l,r,c)\NotDownarrowTwo_{ai}
N''\wedge (l',r,c)\NotDownarrowTwo_{ai} N''}$. Furthermore\\[1.2ex]
\hspace*{4ex}$\mathsf{N'\equiv n[\langle v
\rangle^{\delta}.P]^{c}_{l',r}|\prod\limits_{i\in
I}n_{i}[(x_{i})^{\delta}_{v}.P_{i}]^{c}_{l_{i},r_{i}}|\prod\limits_{j\in
J}n_{j}[P_{j}\{\epsilon/x_{j}\}]^{c}_{l_{j},r_{j}}|\prod\limits_{k\in
K}n_{k}[P_{k}\{\bot/x_{k}\}]^{c}_{l_{k},r_{k}}|N''}$\\[1ex]
\textbf{Proof.} The proof is similar to the one for Lemma 1, and it
uses Lemma 1 itself to handle premises which are input transitions.\hfill $\Box$ \\[2ex]
\textbf{Lemma 3.} If $\mathsf{T\rhd N\xrightarrow{c?v:\delta[l,r]}N'}$, then\\[1.5ex]
\hspace*{4ex}$\mathsf{N\equiv \prod\limits_{i\in
I}n_{i}[in(x_{i}).P_{i}]^{c}_{l_{i},r_{i}}|\prod\limits_{j\in
J}n_{j}[in(x_{j}).P_{j}]^{c}_{l_{j},r_{j}}|\prod\limits_{k\in
K}n_{k}[(x_{k})^{\delta_{k}}_{v_{k}}.P_{k}]^{c}_{l_{k},r_{k}}|N''}$\\[0.8ex]
where $\mathsf{\forall h\in I\cup J\cup K.d(l,l_{h})\leq
r},~~\mathsf{\forall i\in
I.T|_{l_{i},c}=\emptyset},~~\mathsf{\forall j \in J.T|_{l_{j},c}\neq
\emptyset}$ and $\mathsf{(l,r,c)\NotDownarrowTwo_{i}
N''}$. Furthermore\\[1.2ex]
\hspace*{4ex}$\mathsf{N'\equiv \prod\limits_{i\in
I}n_{i}[(x_{i})^{\delta}_{v}.P_{i}]^{c}_{l_{i},r_{i}}|\prod\limits_{j\in
J}n_{j}[in(x_{j}).P_{j}]^{c}_{l_{j},r_{j}}|\prod\limits_{k\in
K}n_{k}[P_{k}\{\bot/x_{k}\}]^{c}_{l_{k},r_{k}}|N''}$\\[0.8ex]
\textbf{Proof.} The proof is similar to the one for Lemma 1, using
rules for begin transmission event.\hfill $\Box$ \\[2ex]
\textbf{Lemma 4.} If $\mathsf{T\rhd N\xrightarrow{c!v:\delta[l,r]}N'}$, then\\[1.5ex]
\hspace*{4ex}$\mathsf{N\equiv n[out\langle
u\rangle.P]^{c}_{l,r}|\prod\limits_{i\in
I}n_{i}[in(x_{i}).P_{i}]^{c}_{l_{i},r_{i}}|\prod\limits_{j\in
J}n_{j}[in(x_{j}).P_{j}]^{c}_{l_{j},r_{j}}|\prod\limits_{k\in
K}n_{k}[(x_{k})^{\delta_{k}}_{v_{k}}.P_{k}]^{c}_{l_{k},r_{k}}|N''}$\\[0.8ex]
where $\mathsf{\forall h\in I\cup J\cup K.d(l,l_{h})\leq
r},~~\mathsf{\forall i\in
I.T|_{l_{i},c}=\emptyset},~~\mathsf{\forall
j \in J.T|_{l_{j},c}\neq \emptyset}$ and $\mathsf{(l,r,c)\NotDownarrowTwo_{i}N''}$. Furthermore if\\[1ex]
$\mathsf{\ThreeQuotersLeft u \ThreeQuotersRight=v}$
and $\mathsf{\Lbag u\Rbag=\delta}$ then\\[1.2ex]
\hspace*{4ex}$\mathsf{N'\equiv n[\langle v
\rangle^{\delta}.P]^{c}_{l,r}|\prod\limits_{i\in
I}n_{i}[(x_{i})^{\delta}_{v}.P_{i}]^{c}_{l_{i},r_{i}}|\prod\limits_{j\in
J}n_{j}[in(x_{j}).P_{j}]^{c}_{l_{j},r_{j}}|\prod\limits_{k\in
K}n_{k}[P_{k}\{\bot/x_{k}\}]^{c}_{l_{k},r_{k}}|N''}$\\[0.8ex]
\textbf{Proof.} The proof is similar to the one for Lemma 2, and it
uses Lemma 3 itself to handle premises which are input transitions.\hfill $\Box$ \\[2ex]
\textbf{Lemma 5.} If $\mathsf{T\rhd N\xrightarrow{\sigma}N'}$, then\\[1.5ex]
\hspace*{4ex}$\mathsf{N\equiv \prod\limits_{i\in I}n_{i}[\langle
v_{i}\rangle^{\delta_{i}}.P_{i}]^{c}_{l_{i},r_{i}}|\prod\limits_{j\in
J}n_{j}[(x_{j})^{\delta_{j}}_{v_{j}}.P_{j}]^{c}_{l_{j},r_{j}}|\prod\limits_{k\in
K}n_{k}[P_{k}]^{c}_{l_{k},r_{k}}}$\\[0.8ex]
where $\mathsf{\forall i\in I.\delta_{i}>0}$ and
$\mathsf{\forall j\in J.\delta_{j}>0}$. Furthermore\\[1.2ex]
\hspace*{4ex}$\mathsf{N'\equiv \prod\limits_{i\in I}n_{i}[\langle
v_{i}\rangle^{\delta_{i}-1}.P_{i}]^{c}_{l_{i},r_{i}}|\prod\limits_{j\in
J}n_{j}[(x_{j})^{\delta_{j}-1}_{v_{j}}.P_{j}]^{c}_{l_{j},r_{j}}|\prod\limits_{k\in
K}n_{k}[P_{k}]^{c}_{l_{k},r_{k}}}$\\[0.8ex]
\textbf{Proof.} The proof is by induction on the size of
$\mathsf{I\cup J\cup K}$, and then for each case it is similar to
the one for Lemma 1, using rules for time passing event. \hfill
$\Box$
\end{document}